\documentclass[12pt,preprint]{aastex}

\slugcomment{Published in the ApJ Letters, vol. 733, L1, 2011 May 20}

\shorttitle{New census of NGC 6791}
\shortauthors{Platais et al.}

\begin{document}


\title{A New Look at the Old Star Cluster NGC~6791}


\author{I. Platais}
\affil{Department of Physics and Astronomy, The Johns Hopkins University,
Baltimore, MD 21218, USA;}
\email{imants@pha.jhu.edu}

\author{K. M. Cudworth}
\affil{Yerkes Observatory, The University of Chicago, Williams Bay, WI 53191,
USA}

\author{V. Kozhurina-Platais}
\affil{Space Telescope Science Institute, 3700 San Martin Drive, Baltimore,
MD 21218, USA}

\author{D. E. McLaughlin}
\affil{Astrophysics Group, Lennard-Jones Laboratories, Keele University,
Keele, Staffordshire ST5 5BG, UK}

\author{S. Meibom}
\affil{Harvard-Smithsonian Center for Astrophysics, 60 Garden Street, Cambridge, MA 02138, USA}

\and

\author{C. Veillet}
\affil{Canada-France-Hawaii Telescope Corporation, Kamuela, HI 96743, USA}


\begin{abstract}
We present comprehensive cluster membership and $g'r'$ photometry of
the prototypical old, metal-rich Galactic star cluster NGC~6791. 
The proper-motion catalog contains 58,901
objects down to $g'\sim$24, limited to a circular area
of radius 30$\arcmin$. The highest precision of the proper motions is
0.08 mas~yr$^{-1}$. 
Our proper motions confirm cluster membership of all main and also some
rare constituents of NGC~6791. The total number of probable cluster members
down to $g'=22$ ($M_V\sim+$8) is $\sim$4800, corresponding to
$M_{\rm tot}\approx5000~M_\odot$. New findings include an extended
horizontal branch in this cluster. The angular radius of NGC~6791 is at least
15$\arcmin$ (the effective radius is $R_h\simeq 4.4\arcmin$ while the 
tidal radius is $r_t\simeq 23\arcmin$). 
The luminosity function of the cluster
peaks at $M_{g'}\sim+$4.5 and then steadily declines toward fainter magnitudes.
Our data provide evidence that differential reddening may not be ignored
in NGC~6791.
\end{abstract}



\section{Introduction}

NGC~6791 is an extreme Galactic star cluster with an old age of
$\sim$8~Gyr \citep{gru08}, a high metallicity [Fe/H]=$+$0.30
\citep{boe09}, and an unusual orbit that periodically brings it close
to the bulge of the Milky Way  \citep{bed06}. \citet{kin65} was the
first to provide an estimate for the total mass of NGC~6791:
$\sim$3700$M_\sun$ down to $V=$20, 
confirmed by \citet{kal92}. It is widely acknowledged that
NGC~6791 is one of the most massive old open clusters in our Galaxy.

As indicated by the discovery of several extremely blue subdwarfs 
\citep{kal92} and the presence of a prominent red clump, morphology
of the color-magnitude diagram (CMD) for NGC~6791 is complex.
The proposed enhanced mass loss along the red giant branch (RGB) 
appears to explain
the presence of hot subdwarfs and the abnormally young 2.4~Gyr white dwarf
cooling age \citep{ka07}.
Recently, \citet{twa11} reported that the CMD in the inner part of
NGC~6791 ($R<2\arcmin$) is somewhat different from that in its outer part
($2\arcmin<R<5\arcmin$). Specifically, there is a dichotomy near
the main-sequence turnoff, which the authors interpret as a
result of protracted star formation.

This cluster is critically important to
understanding a number of issues, such as stellar evolution and
population synthesis at high metallicity in the Milky
Way and other galaxies, specifically, in ellipticals and in spirals
with bulges. Due to its relative proximity and its abundance of
members, NGC~6791 provides perhaps the best opportunity for
detailed studies of old and metal-rich populations.
Despite a rich array of extant photometric and 
spectroscopic studies, however, the basic knowledge of cluster
membership by kinematic means has been limited until now to radial
velocities of a few dozen bright stars 
\citep[e.g.,][]{sco95,car06} 
and preliminary proper motions for a few thousand stars \citep{cud93,bed06}.  
NGC~6791 is located in the field of view (FOV) of NASA's {\it Kepler} mission
\citep{bor10} and is one of the targets of the Kepler Open Cluster
Study \citep{mei10}.
It is also one of the clusters in the WIYN Open Cluster
Study\footnote{This is WOCS paper 46 of the WIYN Open Cluster Study.}
\citep{mat00} -- again, largely owing to its extreme properties.

This Letter presents the highlights of a new and comprehensive proper-motion
study of NGC 6791, based upon the selection of exquisite deep
photographic plates in combination with a decade of CCD imaging. The
resulting proper motions provide nearly definite cluster membership
for stars brighter than $g'\sim$22 and, thus, unveil most of the
cluster's population with unprecedented clarity. 

\section{Astrometry and Photometry}

NGC~6791 is a rich but faint star cluster with a main-sequence turnoff at
$V\sim$17.5 \citep[][hereafter SBG]{ste03}, thus requiring relatively
long exposures even 
with 3--4~m class telescopes. Fortunately, there are deep photographic
plates dating back to 1961. We selected the best 33 plates (assessed
by the depth and sharpness of images) taken with the Lick 3~m and the Kitt
Peak National Observatory (KPNO) 4~m telescopes. These plates were digitised
using the Space Telescope Science Institute's GAMMA~II multi-channel scanning
microdensitometer. The second epoch includes 66 frames obtained with the
NOAO CCD Mosaic Imager at the KPNO 4~m telescope in 1999-2007. The finest
24 CCD mosaic frames in $g'r'$ filters (1, 30, and 300~s exposures)
were obtained in 2009 November with MegaCam at the 3.6~m
Canada-France-Hawaii Telescope. These frames served for construction
of a master catalog consisting of 128,771 objects down to $g'\sim$25
and over an FOV of 1 deg$^{2}$.  We put positions on
the system of the UCAC3 catalog \citep{zac10} and calculated proper
motions on this system using the iterative central-plate-overlap
algorithm \citep[e.g.,][]{jon88}.
The final catalog contains a total of
58,901 objects down to $g'\sim$23.8 over a smaller area of
$\sim$0.8 deg$^2$, defined by a circular FOV (with radius 30\arcmin)
of the KPNO photographic 
plates. The highest precision of our proper motions is 0.08 mas~yr$^{-1}$.
Cluster membership probabilities are calculated using the frequency
functions for cluster and field stars, obtained in selected magnitude
and spatial ranges. Discrimination between cluster members and field
stars is excellent down to $g'\sim$22.  Details of the astrometric
reductions and the description of the catalog itself will appear
elsewhere (Platais et al., in preparation).

For photometry, we used only the MegaCam images in $g'r'$ filters.
The central part of NGC~6791 is fairly crowded, often resulting in
blended images. We note that all MegaCam images are preprocessed
with Elixir which, besides the standard steps, also normalizes the
instrumental photometric zeropoint between the chips
\citep[see][]{mag04,cle08}.
The short, 1~s exposures obtained in subarcsecond
seeing were instrumental in constructing an initial catalog of
positions and aperture photometry (MAG\_AUTO magnitudes from
SExtractor). 
Magnitudes from longer exposures were added only if they matched
specified tolerances on position and magnitude. In this way it was
possible to minimize the deleterious effects of image
blending. Nevertheless, image blending and CCD chip edge effects
may occasionally yield stellar colors that are significantly off.
Our instrumental 
photometry was put into the standard $g'r'$ system by employing the extant
secondary photometric sequence in NGC~6791 \citep{cle08}. The rms error
of the calibrating fits is 0.02 mag.  On average, a calibrated
$g'$ magnitude near the main-sequence turnoff of NGC~6791 is $\sim$0.41~mag
fainter than that in $V$.

\section{New Census of NGC~6791}

One of the most comprehensive photometric surveys of NGC~6791 is that
by SBG, who also provide extensive lists of bright stars,
red giants, and faint blue stars in this cluster---all possible or
likely members. Now, we are in a position to provide nearly definite
star-by-star kinematic membership, with a caveat that stars with
blended images may not appear in our proper motion catalog or, if they
do, may be biased.
An example of such a case is the eclipsing binary V20 --
a definite cluster member \citep{gru08} but incorrectly  classed
as a field star in our catalog. Its measured proper motion is apparently
corrupted by the adjacent star ($\Delta_{\rm pos}$=2$\arcsec$) in the amount
of $\sim$6$\sigma_{\mu}$.
The CMD in Figure~1 shows the cleanest and the
most complete census of NGC~6791 to date. It contains all stars with
cluster membership probability $P_{\mu} \geq$19\%, as well as stars
with $19\% > P_{\mu} > 1$\% but located on the main sequence,
all within a radius $R\leq 15\arcmin$ from the cluster center --
very close to that adopted by \citet{twa11}. The cluster membership
probabilities 
for peripheral members of NGC~6791 ($R>15\arcmin$) are less reliable
because of the overwhelming number of field stars and numerical
instabilities in finding the cluster population.
Hence, in this region the selection of cluster members is limited to
stars that are on the main sequence and have $P_{\mu}>$2\%.
From the combined CMD we deleted all stars below the main sequence and 
with $g'>19$, except for a few very blue stars with
$P_{\mu}\ga 50$\%. Likewise, all stars with $g'>16$ and located well above
the main sequence and to the right of the RGB were
eliminated. This leaves us with a total of 5699 probable cluster
members.

\subsection{Horizontal branch}

The most prominent part of the horizontal branch (HB)---the red
clump---contains 26 now kinematically-confirmed stars, consistent with
the photometric selection 
by \citet{kal92}. More exciting is the presence of extreme
horizontal branch
(EHB) stars in NGC~6791, first recognized by \citet{kal92}.
The following EHB stars are kinematic cluster members: B1,...,B7, B9
\citep[names from][]{kal92,kal95}. A much fainter star, B8, has a marginal
$P_{\mu}$=10\%; however, its astrometry has large uncertainties and
therefore we consider it to be a possible cluster member. In addition,
star B16 ($P_{\mu}$=59\%) appears to be a cluster member and the following SBG
stars are likely cluster members as well: S3472 ($P_{\mu}$=98\%),
 S13881 ($P_{\mu}$=60\%). The brightest
UV-source, B10, \citep{lan98} is a definite field star. In Table~1 we
provide two new EHB stars in NGC~6791, located at $R\sim10\arcmin$.

Besides new EHB candidates, Table~1 lists a number of probable cluster
members that we argue are classical HB stars. The bluest of these is
the well-studied star 2-17 \citep{pet98}, which is frequently assigned
to the EHB. An intriguing feature of the HB here
is an upturn at its red end. This could be the effect of duplicity
among the red clump stars, although that may not be the only
interpretation.
We stress that proving the reality of the HB in NGC~6791 depends
critically upon the ability to distinguish HB stars from
blue stragglers.

There is one very bright star, S12158 ($V=$11.96
$B-V=$1.24), which is formally a probable cluster member 
($P_{\mu}=$91\%) but defies classification. Based on its possibly
abnormally high luminosity, we consider it 
provisionally to be a field star. 

\subsection{Sub-subgiants}

M67 is the first open cluster where the so-called  sub-subgiants were
discovered \citep{mat03}. These apparent binary stars occupy an area
of the CMD below the subgiant branch, which is not easy to populate
with any combination of two normal cluster stars. Owing to the low
frequency of field stars in this part of the CMD, we are confident
that at least five likely sub-subgiants are present in NGC~6791.
In the notation of SBG, these are:
S83, S746, S3626, S13753, and S15561.
Three of these are also the known variable stars V17, V59, and
01431\_10 listed by \citet{mar07}. Only star V17 is in common with the
alternative list of sub-subgiants (red stragglers) given by
\citet{kal03}.

\section{Luminosity function and density profile}

The depth and spatial extent of our survey allow the derivation of a
reliable luminosity function (LF) for the entire cluster down
 to $g'\sim24$. The only omissions are secondaries in binary stars
and severe blends.  We have counted all stars from the 
cropped CMD (Figure~1) in 0.5 mag bins, in two ways:
star-by-star and by weighting each star with its membership
probability (Figure~2). To translate $g'$ into $M_{g^\prime}$,
an apparent distance modulus of 13.56 was applied.
The peak at $M_{g^\prime} \sim +4.5$ and
the smooth downward trend
toward fainter magnitudes is unmistakable. It is consistent with
the appearance of the LF in \citet{kin05}, although we cannot 
replicate the flattening of their LF at $g'>22$---even less so, the
flat luminosity functions provided in \citet{kal92} and \citet{kal95}.
A total of $\sim$1000 cluster members
per 1~mag bin around the main-sequence turnoff
($M_{g^\prime}\sim +4.5$) in our present sample is a record high among the known
old Galactic open clusters. 

We constructed a (number) surface-density profile for NGC 6791 using
the 4830 stars in Figure~1 brighter than $g^\prime=22$ (where 
the proper-motion discrimination between cluster members and field
stars is best). This is
shown in Figure~3. The filled circles in this plot show the number of
stars per unit area, as a function of projected cluster-centric radius,
defined by adding up stars weighted by their membership probabilities in
a series of concentric circular annuli. The open circles
result from adding up the total numbers of the same stars directly,
without weighting by $P_\mu$.

We fitted a single-mass, isotropic \citet{kin66} model to the
$P_\mu$-weighted density profile in Figure~3, using only the data
inside $R<15\arcmin$ to constrain the fit. This implies a rather low
central concentration for the cluster:
$c\equiv\log\,(r_t/r_0) = 0.74\pm0.05$, 
corresponding to a dimensionless central King potential,
$W_0=3.4\pm0.3$. The best-fitting King scale radius is
$r_0=4\farcm22\pm0\farcm30$, which, given the low fitted
concentration, implies a projected core (half-power) radius of
$R_c=3\farcm28\pm0\farcm12$. 
The associated tidal radius is $r_t=23\farcm1\pm1\farcm0$, and the
effective (projected ``half-mass'') radius is
$R_h=4\farcm42\pm0\farcm02$ 
(it is normally the case that estimates of $R_h$ are
more robust than those of other characteristic radii). For a distance
of 4.0~kpc to NGC 6791 \citep{gru08}, $1\arcmin=1.16~{\rm pc}$;
thus, we have that the projected core radius of the cluster is
$R_c\simeq 3.8$~pc; the effective radius is $R_h\simeq
5.1$~pc; and the fitted tidal radius is $r_t\simeq 27$~pc.

Fitting to various other number density profiles defined by both
brighter and fainter magnitude cuts, and both weighted and not
weighted by $P_\mu$, yields results for $R_c$, $R_h$, and $r_t$ that
are generally compatible with the range of values suggested by the
numbers just given for 
the $g^\prime<22$ case. The same is true for fits to luminosity
surface-density profiles defined by adding up the luminosities of all
the stars in our series of annuli. In exploring these various
definitions of the density profile, we saw a clear tendency for
the inferred tidal radius
(and thus the concentration, $\log\,(r_t/r_0)$) to increase
systematically---along with $R_h$, to a lesser extent---as stars with
fainter and fainter magnitudes were included to define the observed
density profile. 
This is primarily because stars with $g^\prime \ga 19$--20 in our
sample are somewhat depleted, relative to the brighter stars, in the
innermost $\sim\!2\arcmin$--$3\arcmin$ of NGC 6791. However, it is not
clear whether this is a physical effect (e.g., due to mass
segregation) or an artifact of crowding and image blending in the core
of the cluster.

The total luminosity of all stars with $g^\prime<22$ in our catalogue
is $L_{\rm tot}\approx 5500$--$6500~L_{\odot,g^{\prime}}$
(assuming $D=4.0$~kpc, corrected for $A_{\rm g^{\prime}}=0.55$~mag of
extinction, and depending on whether or not the luminosities are
weighted by $P_\mu$). Adopting a theoretical $g^\prime$ stellar
mass-luminosity relation from the Padova isochrones \citep{mar08}
implies a total mass of $M_{\rm tot}\approx5000~M_\odot$ for the
observed stars. This is a lower limit to the cluster mass because we
have not attempted corrections for stellar binarity or incompleteness.
We note that, if a \citet{kin66} model with $c=0.74$ and
$R_h=5$~pc accurately describes the internal mass profile of NGC 6791,
then the projected, one-dimensional velocity dispersion averaged
within the effective radius should be
$\langle\sigma\rangle_h \simeq 0.36\,\left(GM_{\rm tot}/R_h\right)^{1/2}
   \approx 0.75~{\rm km~s}^{-1}$.
This corresponds to an intrinsic proper-motion dispersion of
$\sigma_\mu \approx
    0.04\,\left(D/4\,{\rm kpc}\right)^{-1}~{\rm mas~yr}^{-1}$, 
which is much smaller than the formal uncertainties in our proper
motions. 
It is also significantly lower than the radial-velocity dispersion
of $\approx\!2~{\rm km~s}^{-1}$ reported for about a dozen red giants
by \citet{car06}, suggesting either that the \citeauthor{car06}
dispersion may be spuriously high, or that NGC 6791 may not be in
virial equilibrium. Perhaps related to the second option, we note that
a 5000 $M_\odot$ cluster in a Galactic orbit with a pericenter of
$\simeq\!3$~kpc and an apocenter of $\simeq\!10$~kpc \citep{bed06} is
expected to have a tidal radius of
$r_t\approx 13~{\rm pc}\simeq11\arcmin$ at
pericenter, and $r_t\approx28~{\rm pc}\simeq24\arcmin$ at
apocenter---as against a value of $r_t\ga 23\arcmin$ 
suggested by our King-model fitting of NGC 6791.
Deciding the true dynamical state of this cluster will require much
more comprehensive and higher-precision proper-motion and
radial-velocity surveys, to obtain the tightest possible constraints
on its orbit and to delineate accurately its internal kinematics.

\section{Morphology of the red giant branch}

Even a casual inspection of the cluster's CMD reveals an unusually broad
RGB -- on the order of $\Delta(g'-r')\sim$0.1 mag,
while the formal uncertainties in the colors of these stars do
not exceed $\sim$0.02 mag. There is a variety of potential
sources of increased scatter in the CMD, such as instrumental
effects including the limitations of aperture photometry in crowded fields,
presence of field stars, differential reddening, binarity of stars,
metallicity variations, and possible effects on color/luminosity of RGB
stars by variable mass loss suggested by \citet{ka07}.

Our proper-motion-vetted CMD offers a better look at the morphology
of the RGB, which is more sensitive to differential
reddening than the main sequence (apart from its turnoff region).
In addition, our sample covers the largest area ever considered, thus
enhancing the chances of finding the signs of differential reddening.

Unfortunately, we cannot take advantage of significantly reduced reddening
effects in the near infrared. There is a poorly understood scatter in the
$J-K$ color of our RGB stars from either 2MASS or \citet{car05}, which
prevents us from using this color as a nearly reddening-free reference.
Instead, we examined  the morphology of the RGB branch in $g^\prime r^\prime$
CMD using approximately equal numbers of inner and outer probable 
cluster members by dividing the cluster at $R=$5$\arcmin$.
Then, we draw a ridge line for the inner sample, encompassing 
subgiants and RGB stars; see Figure~4. The majority of RGB stars at
$R<5\arcmin$ are bluer than this ridge line but, surprisingly,
nearly all RGB stars at $R>5\arcmin$ are redder
than the ridge line of the inner sample. 
The shift in RGB color between the inner and outer samples is
$\sim0.05$~mag, whereas the age difference of 1~Gyr suggested by
\citet{twa11} would produce only a $\sim$0.01 mag color shift in
$g^\prime-r^\prime$ across the RGB region \citep{gir04}. 
We conjecture that this color shift, at least in part, is due to
differential reddening. Among the effects listed above, it is the only
physical effect known to have an undeniable spatial dependency.

Our data provide evidence that differential reddening may
not be ignored across the face of NGC~6791. Further advancements in our
understanding of such a small effect require accurate near-infrared
photometry and reliable identification of binaries in the upper part of the CMD.

\section{Conclusions}

This study provides a comprehensive and clean list of cluster members
in one of the astrophysically most important Galactic star clusters.
We hope that our findings in NGC~6791 and excellent cluster membership
will stimulate in-depth future studies of individual cluster members. 

\acknowledgments

We thank D. Crawford, J. Cummings, E. Hubbard, T. Kinman, B. Lynds,
S. Majewski, R. Michie (deceased), K. Mighell,  A. Sarajedini, and
M. Sosey, whose time and effort at the telescope made this
project possible. We thank the referee Barbara Anthony-Twarog
for pointing out the limitations of our data. 
This work has been supported in part by NSF grant
AST 09-08114 to JHU (I.P.) and by NASA grant NNX09AH18A 
(The Kepler Open Cluster Study).
Astrophysics at Keele University is supported by an STFC Rolling
Grant. This study is based on observations obtained with 
MegaPrime/MegaCam, a joint project of Canada-France-Hawaii Telescope (CFHT)
and CEA/DAPNIA, at the CFHT which is operated by the NRC of Canada,
the Institute National des Sciences de l'Universe of the CNRS,
and the University of Hawaii.

\begin{deluxetable}{ccccrcl}
\tabletypesize{\scriptsize}

\tablecaption{New EHB candidates and our selection of HB stars}
\tablewidth{0pt}
\tablehead{
\colhead{ID} & \colhead{RA (2000)} & \colhead{Dec (2000)} & 
\colhead{$g'$} & \colhead{$g'-r'$} &
\colhead{$P_{\mu}$ (\%)} & \colhead{Comment\tablenotemark{a}}
}
\startdata
41698 & $290\fdg2076111$ & $37\fdg6103973$ & 20.291 & $-$0.140 & 46 & EHB candidate \\
58783 & 290.4511108 & 37.7326775 & 17.667 & 0.032 & 56 & EHB candidate \\ 
45106 & 290.1308289 & 37.6360130 & 14.999 & 0.508 & 27 & S2101 \\       
52460 & 290.0726929 & 37.6899757 & 15.116 & 0.845 & 42 & S440 \\
58857 & 290.2606201 & 37.7331924 & 15.186 & 0.530 & 99 & 3-22;  RV=$-$2 var\\
60456 & 290.2799683 & 37.7429771 & 15.186 & 1.017 & 99 & 3-27 \\
64067 & 290.1457214 & 37.7645378 & 15.253 & 0.387 & 99 & S2746 \\               
64589 & 290.2497864 & 37.7675743 & 15.155 & 0.278 & 99 & 2-17;  RV=$-$48 \\
65895 & 290.2140503 & 37.7751503 & 14.785 & 1.075 & 99 & NW-20; RV=$-$54 \\
69734 & 290.2836304 & 37.7970619 & 14.568 & 1.096 & 99 & 3-33; RV=$-$20 \\     
69976 & 290.1916809 & 37.7985649 & 14.984 & 1.045 & 95 & 2-45; RV=$-$53 \\   
82982 & 290.3112488 & 37.8854256 & 15.276 & 0.901 & 53 & S14140 \\             
89107 & 290.1376343 & 37.9322052 & 14.215 & 1.101 & 41 & S2382 \\              
93625 & 290.4263306 & 37.9681969 & 15.165 & 0.590 & 58 &   \\                   
\enddata
\tablenotetext{a}{The SBG numbers are preceded by a letter S; remaining
cross-identifications are from \citet{kin65}; radial velocity (RV) in
km~s$^{-1}$ is from the literature.}
\end{deluxetable}

\clearpage

\begin{figure}
\epsscale{.65}
\plotone{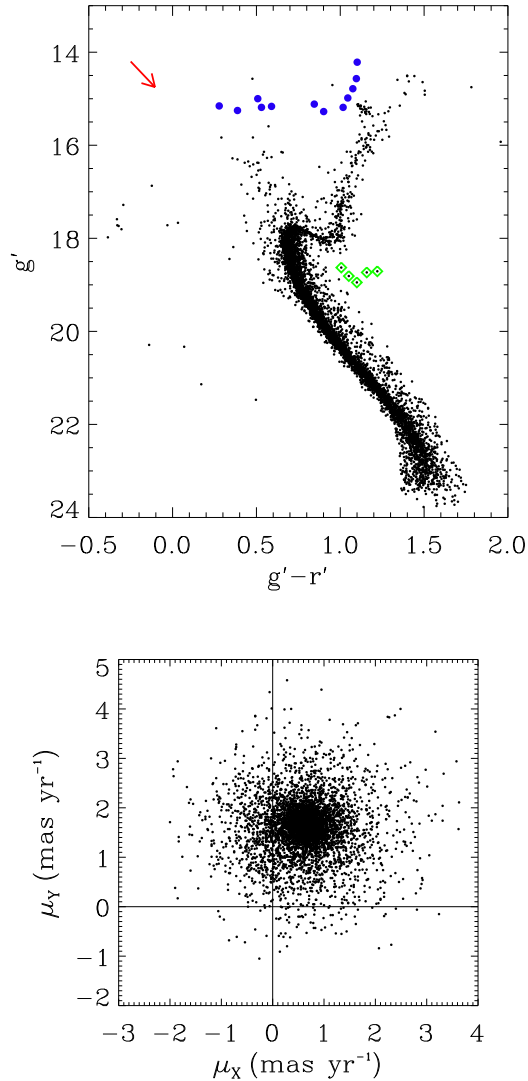}
\caption{Color-magnitude and vector-point diagrams of NGC~6791.
Upper panel: CMD of all probable cluster members, with the exception of
some cropped areas. The large dots indicate our selection of HB stars;
the rombs show the proposed sub-subgiants. The arrow shows the direction and
the adopted total amount of reddening [$E(g'-r')=0.145$, $A_{g'}=0.55$,
following  \citet{gru08}]. 
Lower panel: relative proper motions of probable cluster members only.
The larger spread of the outer points reflects the relatively low
precision of proper motions near the limit of the catalog at $g'=23.5$.
\label{fig1}}
\end{figure}

\clearpage

\begin{figure}
\includegraphics[angle=270,scale=.50]{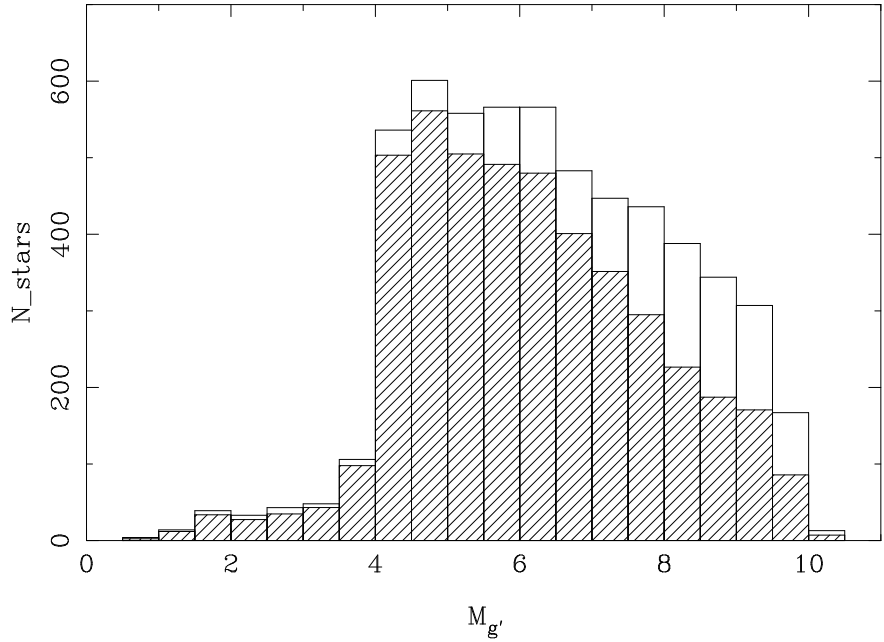}
\caption{Luminosity function of NGC~6791. Each bin is 0.5 mag wide.
The hatched histogram is obtained by weighting each star, shown in
Figure~1, with its membership probability ($P_\mu$), while
the open histogram shows the the actual numbers of stars without
weighting by $P_\mu$. The last two bins at absolute magnitude $M_{g'}\sim+10$ are incomplete.
\label{fig2}}
\end{figure}

\clearpage

\begin{figure}
\epsscale{0.85}
\plotone{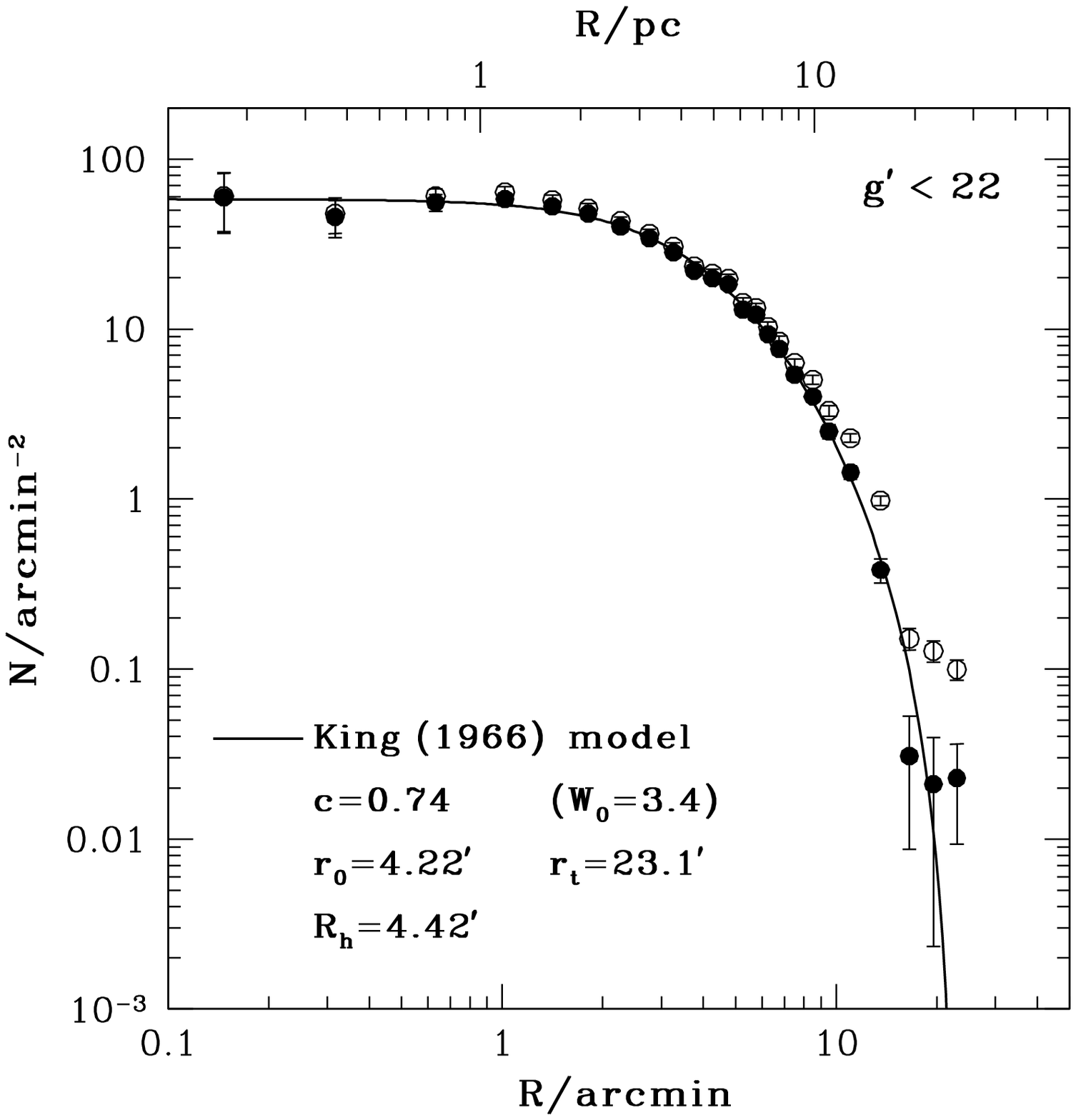}
\caption{Number-density profile of stars in NGC 6791, down to apparent
$g^\prime=22$. Filled circles are the densities obtained by weighting
each star with its  $P_\mu$ in concentric circular annuli;
open circles are the densities obtained by direct adding up the number
of stars.
The curve is the best fit of a single-mass, isotropic \citet{kin66} model
to the $P_\mu$-weighted data inside $R<15\arcmin$. For a distance of
 4.0~kpc to the cluster, $1\arcmin=1.16~{\rm pc}$. 
\label{fig3}}
\end{figure}

\clearpage

\begin{figure}
\epsscale{0.85}
\plotone{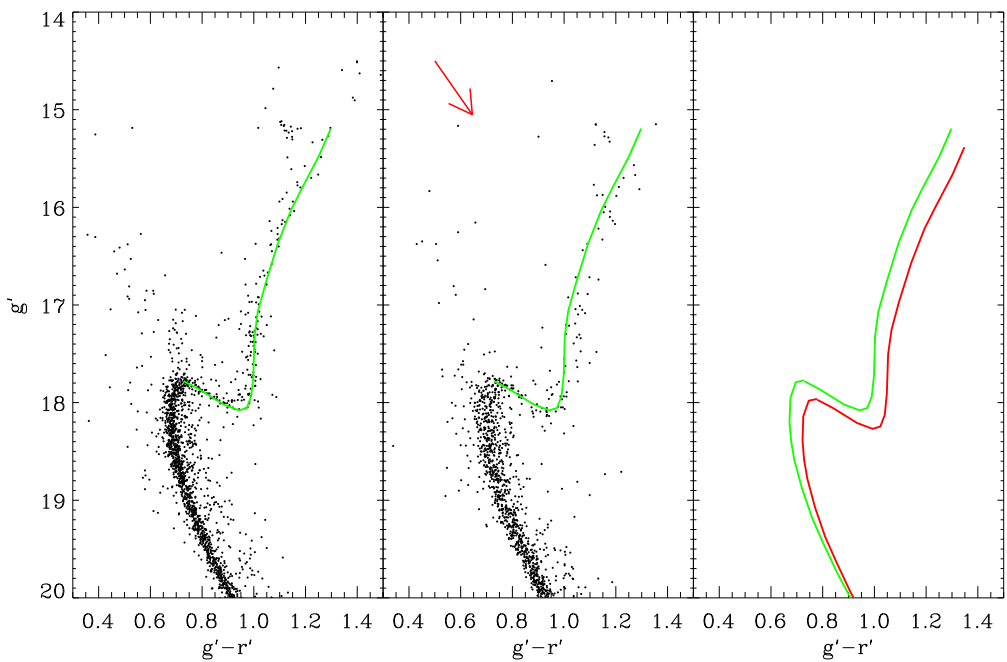}
\caption{Color-magnitude diagram of NGC~6791 with emphasis on RGB.
The left panel shows a CMD for the inner cluster sample
($R<5\arcmin$); the middle panel, for the outer sample
($R\ge5\arcmin$).  For the arrow, see Figure~1.
In both panels, the curve indicates the location of
inner sample's ridge line for subgiants and RGB stars. Note the
location of RGB stars relative to this ridge line. The right panel
elucidates the effect of a reddening $E(g'-r')=0.05$ applied to the cluster's
inner-sample fiducial. This loosely characterizes the degree of
differential reddening in NGC~6791.}  
\label{fig4}
\end{figure}

\end{document}